\documentclass[aps,prd,twocolumn,showpacs,draft,floatfix]{revtex4}
\begin{document}

\title{Noncommutative spaces,
 the quantum of time and the Lorentz symmetry}
\author{Juan~M. Romero}
\email{sanpedro@nucleares.unam.mx} \affiliation{Instituto de
Ciencias Nucleares, Universidad Nacional Aut\'onoma de M\'exico,
Apartado Postal 70-543, M\'exico 04510 DF, M\'exico}
\author{J.~D.~Vergara}
\email{vergara@nucleares.unam.mx} \affiliation{Instituto de Ciencias
Nucleares, Universidad Nacional Aut\'onoma de M\'exico, Apartado
Postal 70-543, M\'exico 04510 DF, M\'exico}
\author{J.~A.~Santiago}
\email{sgarciaj@uaeh.reduaeh.mx}
\affiliation{Centro de Investigaci\'on Avanzada en Ingenier\'\i a
Industrial, Universidad Aut\'onoma del Estado de Hidalgo,
Pachuca 42090, M\'exico}

\date{\today}

\begin{abstract}
We introduce three space-times that are discrete in time and
compatible with the Lorentz symmetry. We show that these spaces are
no commutative, with commutation relations similar to the relations
of the Snyder and Yang spaces. Furthermore, using a reparametrized
relativistic particle  we obtain a realization of the Snyder type
spaces and we construct an action for them.

\end{abstract}


\maketitle

\section{Introduction}
\label{s:intro} Because of several interesting results in string
theory \cite{witten:gnus}, noncommutative spaces have recently
attracted attention. In this context, the noncommutativity parameter
is a constant; which leads to inconsistencies with the usual Lorentz
symmetry \cite{carroll:gnus}. In that case a deformed symmetry,
namely the so-called twisted Poincar\'e invariance holds instead
  \cite{chaichian:gnus}. It should be stressed,
however, that not all proposals of noncommutative spaces are incompatible
with the Lorentz symmetry. In a remarkable work, H.~Snyder constructed a
noncommutative space-time compatible with the Lorentz symmetry which is
discrete in the spatial coordinates \cite{sn:gnus}. Moreover,
G.'t Hooft showed that by considering quantum gravity in $(2+1)$ dimensions
a Snyder-like space-time can be obtained \cite{hooft:gnus}.
This result suggests that quantum gravity in other dimensions may
imply a noncommutative space of this kind.
Another feature that makes the Snyder space-time
(SST) interesting is the fact that it can be mapped to the $k$-Minkowski
space-time \cite{kowalsky:gnus}; which is the arena for the so-called Doubly
Special Relativity theory. It has been shown \cite{freidel:gnus} that this
theory coincides in some aspects with quantum gravity in $(2+1)$ dimensions.
All these features make it worth studying Snyder kind
noncommutative space-times and their realizations.\\

It is worth to mention that C.N. Yang \cite{yang:gnus} based in the
Snyder construction obtain another space-time that is compatible
with the Lorentz symmetry and also it is invariant under
translations. This space-time is discrete in the spatial
coordinates. \\

In this work we construct three space-times that are discrete in
time and compatibles with the Lorentz symmetry, and we show that
these models are noncommutative. The first space has commutation
relations that looks very similar to the SST, the others
 follow a similar pattern to the Yang space. In the case
of the Snyder type space we are able to obtain a realization of the
model using the action of a reparametrized relativistic particle,
and from this result we construct a general action for a particle in
this kind of noncommutative spaces. We remark that this result is
applicable no only to spaces that are discrete in time, since is
valid also for Snyder like theories with discrete space. \\

The manuscript is organized as follows: In section \ref{s:quan} we
construct the proposed models and show that are noncommutative.
Section \ref{s:real} is composed of three parts. In subsection
\ref{ss:par} we review the parametrized free particle, whereas in
subsection \ref{ss:snyder} we provide a realization of the Snyder
type spaces. Furthermore, a general form for the action of a
particle in a Snyder-like noncommutative space with arbitrary
Hamiltonian is presented in subsection \ref{ss:cond}. Finally, we
summarize our results in section \ref{s:summ}.

\section{The quantum of time}
\label{s:quan}

We construct here three examples of  space-times that are discrete
in time and compatible with the Lorentz symmetry. First we introduce
the space with commutation relations similar to the SST and
following a similar approach we obtain two spaces that resemblance
the Yang space.
\subsection{Noncommutative Snyder-like space-time}

Let us start by considering a $(D+2)$-dimensional
space with $\zeta^{A}=\left(\zeta^{0},\zeta^{0^{\prime}},\zeta^{1},
\cdots,\zeta^{D}\right)$ a vector in this space and a metric $\eta$ with
components given by
\begin{eqnarray}
\eta_{00}&\!=\!&1,\qquad\ \ \ \ \,\eta_{0^{\prime} 0^{\prime}}=1,\nonumber\\
\quad \eta_{0 0^{\prime}}&\!=\!&\eta_{0^{\prime} 0}=0,
\quad \eta_{i0^{\prime}}=\eta_{0^{\prime} i}=0,\nonumber\\
\eta_{i0}&\!=\!&\eta_{0i}=0,\quad\ \
\eta_{ij}=-\delta_{ij}, \quad i,j=1,\cdots,D.\qquad
\label{eq:snydermetrica}
\end{eqnarray}
If the transformations $\Lambda$ let the quadratic form
$\tilde S^{2}=\zeta^{T}\eta\zeta=-(\zeta^{0^{\prime}})^{2}-(\zeta^{0})^{2}
+ (\zeta^{1})^{2}+\cdots+(\zeta^{D})^{2}$ invariant, then they satisfy
the identity
\begin{equation}
\Lambda^{T}\eta\Lambda =\eta.\label{eq:translo}
\end{equation}
For the infinitesimal transformations close to the identity,
$\Lambda=I+\epsilon M$, with $I$ the identity matrix, Eq.
(\ref{eq:translo}) implies
\begin{equation}
(I+\epsilon M)^{T}\eta(I+ \epsilon M)=\eta,
\end{equation}
and thus $M^{T}\eta =-\eta M$. That is
\begin{equation}
M_{AB}=-M_{BA}. \label{eq:transsnyder}
\end{equation}
From this matrix we define the infinitesimal transformation
$\delta\zeta^{A}=\epsilon M^{A}\-_{\!B}\zeta^{B}$ with the generators of
the group given by
\begin{eqnarray}
&V\!\!&=\epsilon M^{A}\-_{\!B}\zeta^{B}\frac{\partial }{\partial\zeta^{A}}
= \epsilon M_{AB}\zeta^{B}\frac{\partial }{\partial\zeta_{A}}\nonumber\\
& &=\epsilon M_{00^{\prime}}\left(\zeta^{0^{\prime}}
\frac{\partial }{\partial\zeta_{0}} -
\zeta^{0}
\frac{\partial }{\partial\zeta_{0^{\prime}}}\right)\nonumber\\
 & &{}+\epsilon M_{0^{\prime}i}\left(\zeta^{i}
\frac{\partial }{\partial\zeta_{0^{\prime}}}
-\zeta^{0^{\prime}}\frac{\partial }{\partial\zeta_{i}}\right)\nonumber\\
& &{}+\epsilon M_{0i}\left(\zeta^{i} \frac{\partial }{\partial\zeta_{0}}
-\zeta^{0}\frac{\partial }{\partial\zeta_{i}}\right)\nonumber\\
 & &{}+\epsilon
\frac{1}{2}M_{ji}\left(\zeta^{i} \frac{\partial }{\partial\zeta_{j}}
-\zeta^{j}\frac{\partial }{\partial\zeta_{i}} \right).
\end{eqnarray}
Then, we define the operators
\begin{eqnarray}
l^{00^{\prime}}&=&\zeta^{0^{\prime}} \label{eq:L00}
\frac{\partial }{\partial\zeta_{0}} -\zeta^{0}
\frac{\partial }{\partial\zeta_{0^{\prime}}}, \label{eq:L00'}\\
l^{i0^{\prime}}& =&\zeta^{0^{\prime}}\frac{\partial }{\partial\zeta_{i}}
-\zeta^{i}
\frac{\partial }{\partial\zeta_{0^{\prime}}}, \label{eq:Li0'}\\
l^{0i}&=& \zeta^{i} \frac{\partial }{\partial\zeta_{0}}-
\zeta^{0} \frac{\partial }{\partial\zeta_{i}}, \label{eq:L0i}\\
 l^{ji}&=&\zeta^{i} \frac{\partial }{\partial\zeta_{j}}-
\zeta^{j} \frac{\partial }{\partial\zeta_{i}}. \label{eq:Lji}
\end{eqnarray}
We go now onto considering the reduced space of dimension $(D+1)$,
$\zeta^{\mu}=(\zeta^{0},\zeta^{1},\cdots,\zeta^{D})$ a vector in this
space and the Lorentz transformation in it. As the variable
$\zeta^{0^{\prime}}$ is invariant under the Lorentz transformation in this
reduced space, then $R^{\mu}=l^{\mu 0^{\prime}}$ is a contravariant vector.

Therefore, we define the Hermitian operator
\begin{equation}
\hat X^{\mu}=-ia\left(\zeta^{0^{\prime}}
\frac{\partial }{\partial\zeta_{\mu}}-
\zeta^{\mu} \frac{\partial }{\partial\zeta_{0^{\prime}}}\right),
\quad \mu=0,1,\cdots,D.\quad \label{eq:Xmu}
\end{equation}
where $a$ is a constant of unit length. By employing this, we construct
the Hermitian operator invariant under the Lorentz transformation
\begin{equation}
\hat S^{2}= \hat X_{\mu} \hat X^{\mu}.
\end{equation}
From (\ref{eq:Xmu}) we find the commutation rules
\begin{eqnarray}
\big[\hat X_{\mu} ,\hat X_{\nu}\big] &=&\frac{i a^{2}}{\hbar}
\hat L_{\mu\nu},
\quad 
\label{eq:1}\\
\big[\hat X_{\mu} ,\hat P_{\nu}\big] &=&i\hbar\left(\eta_{\mu\nu}+
\frac{a^{2}}{\hbar^{2}} \hat P_{\mu}\hat P_{\nu} \right),
\label{eq:snydercom}\\
\big[\hat P_{\mu} ,\hat P_{\nu}\big] &=&0, \label{eq:3}
\end{eqnarray}
where
\begin{eqnarray}
\hat L_{\mu\nu}&=&\hat X_{\mu} \hat P_{\nu}- \hat X_{\nu}\hat P_{\mu},
\label{eq:snydermomang}\\
\hat P_{\mu}&=&\frac{-\hbar}{a}\frac{\zeta_{\mu}}{\zeta^{0^{\prime}}}.
\label{eq:snydermom}
\end{eqnarray}
Therefore, this space is noncommutative and has the Lorentz symmetry.\\

Now, it can be shown that $\psi_{i}=e^{-i\tilde L\varphi_{i}}$, with
$\varphi_{i}=\hbox{arctanh}({\zeta^{i}}/{\zeta^{0^{\prime}}})$, is
eigenfunction of $\hat X^{i}$,
\begin{equation}
\hat X^{i}\psi_{i}=a\tilde L\psi_{i}.
\end{equation}
In this case $\tilde L$ can take any arbitrary value. Analogously
$\psi_{0}=e^{i L\varphi_{0}}$, but now with
$\varphi_{0}=\arctan({\zeta^{0}}/{\zeta^{0^{\prime}}})$, is
eigenfunction of $\hat X^{0}$,
\begin{equation}
\hat X^{0}\psi_{0}=aL\psi_{0}.
\end{equation}
As the tangent is $2\pi$-periodic, $\varphi_{0}+2\pi=
\arctan({\zeta^{0}}/{\zeta^{0^{\prime}}})$. Therefore, in order to avoid
$\psi_{0}$ from being a multivalued function, one must constrain the
values of $L$ to be integers. Thus, the time is quantized:
\begin{equation}
t_{N}= N\frac{a}{c},
\end{equation}
with $N$ an integer and $c$ the speed of light. Thus, this
space-time is discrete in time and consistent with the Lorentz
symmetry. Notice that, the Lorentz symmetry is generated by
(\ref{eq:snydermomang}) and that these generators are included in
the generators of the conformal group $SO(D,2)$
(\ref{eq:L00})-(\ref{eq:Lji}). So, the Lorentz symmetry of this
space is a contraction of the conformal group.\\

Notice that we can also take the reduced space
$\zeta^{\alpha}=(\zeta^{0},\zeta^{1},\cdots,\zeta^{d}), \quad d=D-1$
and define $V^{\alpha}= l^{\alpha D},\quad (\alpha=0,1,\cdots,d\,).$
In the $(d+1)$-dimensional Minkowski space, where $\zeta^{D}$ is an
invariant and $V^{\alpha}$ is a contravariant vector. So, we can
define the Hermitian operator
\begin{equation}
\hat X^{\alpha}=-ia\left(\zeta^{D} \frac{\partial }{\partial\zeta_{\alpha}}
-\zeta^{\alpha} \frac{\partial }{\partial\zeta_{D}}\right).
\end{equation}
This case yields the commutation rules
\begin{eqnarray}
\big[\hat X_{\mu} ,\hat X_{\nu}\big] &=&\frac{-i a^{2}}{\hbar}
\hat L_{\mu\nu},
\label{eq:1s}\\
\big[\hat X_{\mu} ,\hat P_{\nu}\big] &=&i\hbar\left(\eta_{\mu\nu}-
\frac{a^{2}}{\hbar^{2}} \hat P_{\mu}\hat P_{\nu} \right),\\
\big[\hat P_{\mu} ,\hat P_{\nu}\big] &=&0, \label{eq:3s}
\end{eqnarray}
which are those of the SST \cite{sn:gnus}. The difference with the space
we constructed here is that in the SST the time is continuous and the
spatial coordinates are discrete.\\

\subsection{Noncommutative Yang-like Space-times}

The Snyder space-time lacks translational invariance, implying in
principle the nonconservation of the energy-momentum in field
theory. Based in this observation Yang \cite{yang:gnus} proposed
another version of a noncommutative space-time. This space is
invariant under Lorentz transformations and infinitesimal
translations. The Yang's construction is similar to the Snyder's,
but requires an extra dimension. In this subsection we modify the
Yang's construction to consider space-times
that are continuous in space but discrete in time. \\

Assume that we have a flat space-time of dimension $D+3$ with
coordinates  $(\zeta^{0},\zeta^{1},...,\zeta^{D}, \zeta^{r},
\zeta^{r^{\prime}}).$ We define the labels $s,s^{\prime},$ with
$s=0$ ($s^{\prime}=0$) iff $\zeta^{r}$ ($\zeta^{r^{\prime}}$) is a
temporal coordinate and $s=1$ ($s^{\prime}=1$) iff $\zeta^{r}$
($\zeta^{r^{\prime}}$) is a spatial coordinate. In this context, the
Hermitic differential operators,
\begin{eqnarray}
\hat X^{\mu}&=&-ia\left(\zeta^{r}
\frac{\partial }{\partial\zeta_{\mu}}-
\zeta^{\mu} \frac{\partial }{\partial\zeta_{r}}\right),
\quad \mu=0,1,\cdots,D, \\
\hat P^{\mu}&=&-i\frac{\hbar}{b}\left(\zeta^{r^{\prime}}
\frac{\partial }{\partial\zeta_{\mu}}-
\zeta^{\mu} \frac{\partial }{\partial\zeta_{r^{\prime}}}\right),
\end{eqnarray}
are contravariant vectors under the transformations that leave
invariant the components $\zeta^{s}, \zeta^{s^{\prime}}$. Using
these operators we obtain the algebra,
\begin{eqnarray}
\big[\hat X^{\mu},\hat X^{\nu}\big]&=&i\frac{a^{2}}{\hbar}(-)^{s}
\hat l^{\mu\nu}
\label{eq:yang1} \\
\big[\hat X^{\mu},\hat P^{\nu}\big]&=&\frac{i\hbar\hat \epsilon}{b}
\eta^{\mu\nu},\\
\big[\hat P^{\mu},\hat P^{\nu}\big]&=&i\frac{\hbar}{b^{2}}
(-)^{s^{\prime}}\hat l^{\mu\nu}, \label{eq:yang2}
\end{eqnarray}
where,
\begin{eqnarray}
\hat l^{\mu\nu}&=&-i\hbar \left(\zeta^{\mu}
\frac{\partial }{\partial\zeta_{\nu}}-
\zeta^{\mu} \frac{\partial }{\partial\zeta_{\nu}}\right),\\
\hat \epsilon&=& -ia\left(\zeta^{r}
\frac{\partial }{\partial\zeta_{r^{\prime}}}-
\zeta^{r^{\prime}} \frac{\partial }{\partial\zeta_{r}}\right).
\end{eqnarray}
By considering the operator $\hat L_{\mu\nu}$ from
(\ref{eq:snydermomang}), the commutation rules can be written as
\begin{eqnarray}
\hat L^{\mu\nu}&=& \frac{\hat \epsilon}{b}l^{\mu\nu}=
\frac{i(-)^{s+1}\hbar \hat \epsilon}{a^{2}b}
\big[\hat X^{\mu}, \hat X^{\nu}\big]\nonumber\\
&=&\frac{i(-)^{s^{\prime+1}} b\hat\epsilon}{\hbar}
(-)^{s^{\prime}}\big[\hat P^{\mu}, \hat P^{\nu}\big],
\end{eqnarray}
and also we have
\begin{eqnarray}
\big[\hat X^{\mu}, \hat \epsilon\big]
&=&i\frac{a^{2}b}{\hbar}(-)^{s+1}\hat P^{\mu}, \label{eq:yang3}\\
\big[\hat P^{\mu}, \hat \epsilon\big]&=&
i\frac{\hbar}{b}(-)^{s^{\prime}} \hat X^{\mu},\\
\big[\hat X^{\beta}, \hat l^{\mu\nu}\big]
&=&i\hbar\left(\hat X^{\mu}\eta^{\beta\nu}-\hat X^{\nu}\eta^{\beta\mu}\right),
\label{eq:yang4a}\\
\big[\hat P^{\beta}, \hat l^{\mu\nu}\big]
&=&i\hbar\left(\hat P^{\mu}\eta^{\beta\nu}-\hat P^{\nu}\eta^{\beta\mu}\right).
\label{eq:yang4}
\end{eqnarray}
 Clearly  the commutation relations
(\ref{eq:yang1})-(\ref{eq:yang2}) are compatible with the Lorentz
symmetry and from (\ref{eq:yang4a})-(\ref{eq:yang4}) we
observe that $\hat l^{\mu\nu}$ is the generator of this group.
Now, taking into account that the translation operator is given by
\begin{eqnarray}
U(\alpha)=e^{-i\alpha_{\mu} \hat P^{\mu}}\approx 1-i\alpha_{\mu}
\hat P^{\mu},\qquad \alpha_{\mu}={\rm const},
\end{eqnarray}
and using the Eqs. (\ref{eq:yang3})-(\ref{eq:yang4}), we get the
transformation rules for the operators
\begin{eqnarray}
U^{-1}(\alpha)\hat X^{\mu}U(\alpha)&\approx&\hat X^{\mu}+\frac{\hbar}{b}
\alpha^{\mu}\hat \epsilon,\\
U^{-1}(\alpha)\hat P^{\mu}U(\alpha)&\approx&\hat P^{\mu}+\frac{\hbar}{b^{2}}
(-)^{s^{\prime}+1}\alpha_{\nu}\hat l^{\nu\mu},\\
U^{-1}(\alpha)\hat l^{\mu\nu} U(\alpha)&\approx& \hat
l^{\mu\nu}+\hbar \left(\alpha^{\mu}\hat P^{\nu}-\alpha^{\nu} \hat
P^{\mu}\right).
\end{eqnarray}
Applying these transformation rules we can show that the commutation
relations (\ref{eq:yang1})-(\ref{eq:yang2}) are invariant under
translations. In consequence for each value of $s$ and $s^{\prime},$
we get a noncommutative space compatible the translations and
Lorentz symmetries. These space-times are discrete in time or in the
spatial coordinates. For example, in the case that
$s=1,s^{\prime}=1,$ i.e. when both extra coordinates are spatial we
get the usual Yang space \cite{yang:gnus}. In this case the Lorentz
group is obtained as a subgroup of $SO(D+2,1)$. Whereas, if
$s=1,s^{\prime}=0,$ we get also discrete spatial coordinates. For
this case the Lorentz group corresponds to a subgroup of
$SO(D+1,2),$ this space was found in \cite{tanaka:gnus}. Now, for
$s=0,s^{\prime}=1$ the temporal coordinate is discrete and the
spatial ones are continuous, for this situation the Lorentz group is
obtained as a subgroup of $SO(D+1,2).$ The last case corresponds to
$s=0,s^{\prime}=0$, here the temporal coordinate is discrete and the
spatial coordinates are continuous, the Lorentz group is in this
case a subgroup of
$SO(D,3).$\\

Other discrete-time models can be found in
\cite{bojowald:gnus,hooft1:gnus, matschull:gnus}.
 Reference \cite{matschull:gnus} is
particularly remarkable as from quantum gravity in $(2+1)$ dimensions
the authors obtain a momenta space having two time-coordinates. The
spectrum of the space-time is similar to the one here obtained, but the
commutation rules are not the same. It is worth mentioning that it was
shown recently that different physical systems can be unified by a
two time-coordinates model \cite{bars:gnus}. A proposal with two
time-coordinates at the level of string theory can be seen in
\cite{vafa:gnus}.\\

In the literature exist some proposals where are analyzed
space-times with noncommutativity in time, these spaces present to
the level of field theory problems with causality and unitarity
\cite{seiberg-chai:gnus}. However, in these examples the Lorentz
symmetry is broken, or they have a twisted Poincar\'e symmetry.
Furthermore, in all these cases the parameter of noncommutativity is
a constant, implying a noncommutative product of Moyal type. Whereas
in our case we have a nonconstant noncommutative parameter implying
a Konsevich product \cite{kontsevich:gnus} and in consequence the
above results are not directly applicable to our spaces.\\

Next we go on to constructing an explicit realization of the
space-time having the commutation rules (\ref{eq:1})--(\ref{eq:3}).

\section{Realizations of noncommutative spaces}
\label{s:real}

A way to obtain realizations of noncommutative spaces comes from mechanical
systems \cite{stern:gnus}. In particular, several authors have recently
reported realizations of the SST; one of them within the
so-called two-times physics \cite{nos:gnus}. Other realizations have been
obtained by considering the dynamics of a free particle; and remarkable
references on that can be found in \cite{girelli:gnus}. It is worth pointing
out that in the realizations based on the free particle, the parameter for
noncommutativity depends on the particle mass as $\theta\sim 1/m$, and thus
the model loses meaning for massless particles.\\

From a parametrized relativistic particle we obtain in this section
a realization of Snyder-like noncommutative spaces; i.e. spaces
having the commutation rules (\ref{eq:1})--(\ref{eq:3}) or
(\ref{eq:1s})--(\ref{eq:3s}). We show, in addition, that such a
realization remain meaningful even for massless particles. A general
form for the action of a particle in this
kind of space-time having an arbitrary Hamiltonian is also provided.\\

\subsection{Parametrized relativistic particle}
\label{ss:par}
In this part we briefly review the parametrized relativistic particle.
We start by showing that an action of the form
\begin{equation}
S= K\int d\tau \sqrt{\cal L}, \qquad K={\rm const}, \label{eq:mara}
\end{equation}
is equivalent to
\begin{equation}
S= \frac{1}{2}\int d\tau \left[\frac{\cal L}{\lambda}+ \lambda K^{2}\right].
\label{eq:maniana}
\end{equation}
This can be seen by obtaining the equation of motion for $\lambda$ from
action (\ref{eq:maniana}),
\begin{equation}
\lambda =\frac{\sqrt{\cal L}}{K},
\end{equation}
and then substituting it back into (\ref{eq:maniana}) to directly obtain
(\ref{eq:mara}). Notice, however, that the $K=0$ case can be considered
from action (\ref{eq:maniana}) but not from (\ref{eq:mara}).\\

Now, the action of the free particle is
\begin{equation}
S=-mc\int^{\tau_{2}}_{\tau_{1}} d\tau\sqrt{\dot X^{\mu}\dot X_{\mu}},
\label{eq:eins1}
\end{equation}
which is, then, equivalent to
\begin{equation}
S_{*}=\frac{1}{2}\int^{\tau_{2}}_{\tau_{1}} d\tau\left[
\frac{\dot X^{2}}{\lambda}+\lambda m^{2}c^{2}\right]\label{eq:eins3}.
\end{equation}
In this case $K=-mc$, so for $m=0$ we are dealing with a massless particle
such as the photon. Some relevant applications of action (\ref{eq:eins3})
at the level of field theory can be found in \cite{strassler:gnus}.\\

The equations of motion for action (\ref{eq:eins3}) are
\begin{eqnarray}
\frac{d}{d\tau}\left(\frac{\dot X^{\mu}}{\lambda}\right)&=&0,
\label{eq:ecuacioneqv}\\
-\frac{\dot X^{2} }{\lambda^{2}}+m^{2}c^{2}&=&0. \label{eq:ecuacioneqv2}
\end{eqnarray}
From the second we obtain
\begin{equation}
\lambda=\frac{\sqrt{\dot X^{2}} }{mc}.
\end{equation}
Thus, by taking $\tau=\tau_{p}$ with $\tau_{p}$ being the proper time,
one gets to
\begin{equation}
\lambda=\frac{1}{m}.
\end{equation}
In this case action (\ref{eq:eins3}) becomes
\begin{equation}
S_{*}=\frac{1}{2}\int^{\tau_{2}}_{\tau_{1}} d\tau\left(m
\dot X^{2}
+mc^{2}\right).\label{eq:accieq1}
\end{equation}
If $m=0$ no definition of proper time exists. However, we can take the
condition $\lambda=1/m_{\nu}$, with $m_{\nu}=h\nu/c^{2}$ the equations
of motion (\ref{eq:ecuacioneqv})--(\ref{eq:ecuacioneqv2}) become
\begin{eqnarray}
\frac{d^{2} X^{\mu}}{ds^{2}}&=&0,
\nonumber\\
\dot X^{2} &=&0,
\end{eqnarray}
which are consistent with the equations of motion of a massless free
particle. For this case action (\ref{eq:eins3}) takes the form
\begin{equation}
S_{*}=\frac{m_{\nu}}{2}\int^{\tau_{2}}_{\tau_{1}}
d\tau\left(\dot X^{2}\right). \label{eq:accieq2}
\end{equation}
Now, by defining  $m_{\gamma}=m$ if $m\not =0$ and $m_{\gamma}=m_{\nu}$
if $m=0$, actions (\ref{eq:accieq1}) and (\ref{eq:accieq2}) can be
rewritten as
\begin{equation}
S_{*}=\frac{1}{2} \int^{\tau_{2}}_{\tau_{1}}d\tau
\left(m_{\gamma}\dot X^{2} +mc^{2}\right).
\label{eq:accred}
\end{equation}
We point out that, contrary to action (\ref{eq:eins1}), action
(\ref{eq:accred}) is not invariant under reparametrizations. If we
want (\ref{eq:accred}) to have this invariance, we must introduce
an extra parameter $\zeta$; which we assume a relativistic invariant.
Thus, the action invariant under reparametrizations is
\begin{equation}
S=\frac{1}{2} \int^{\tau_{2}}_{\tau_{1}}
d\tau\left(\frac{m_{\gamma}\dot X^{2}}{\dot \zeta} +mc^{2}\dot \zeta\right).
\label{eq:accredrep}
\end{equation}
This action is a generalization of the relativistic particle, it is
clear that only for the case $\dot \zeta=1$ we recover the usual
case on the proper time gauge. To analyze the difference between
this system and the usual relativistic particle, we consider the
Hamiltonian analysis of the action (\ref{eq:accredrep}). The
canonical momenta that one obtains from (\ref{eq:accredrep}) are
\begin{eqnarray}
P_{\mu}&=&m_{\gamma}\frac{\dot X_{\mu}}{\dot \zeta},\label{eq:luis0}\\
P_{\zeta}&=&\frac{1}{2}\left(-m_{\gamma}\frac{\dot X_{\mu}\dot X^{\mu}}
{\dot\zeta ^{2}}+mc^{2}\right),
\label{eq:luis}
\end{eqnarray}
and the equations of motion can be written as
\begin{equation}
\dot P_{\mu}=0, \qquad \dot P_{\zeta}=0.\label{eq:eq}
\end{equation}
Now, from the canonical momenta (\ref{eq:luis0}) and (\ref{eq:luis}) one
obtains the constraint
\begin{equation}
\phi =P_{\zeta}+\frac{1}{2m_{\gamma}}
\left(P_{\mu}P^{\mu}-m^{2}c^{2}\right)\approx 0,
\label{eq:poper}
\end{equation}
which implies that, if $P_{\zeta}\not= 0$ then
$P_{\mu}P^{\mu}-m^{2}c^{2}\not =0$. That is, the dispersion relation is
changed. Moreover, by taking $\tau =\tau_{p}$, with $\tau_{p}$ being
the proper time, from (\ref{eq:poper}) one gets to
\begin{equation}
\tau_{p}=\zeta
\left(1- \frac{2}{m_{\gamma}c^{2}}P_{\zeta}\right)^{\frac{1}{2}}.
\label{eq:fundamental}
\end{equation}
Therefore, there exists a relation between $\zeta$ and $\tau_{p}$.
Notice that this relationship implies
\begin{equation}
P_{\zeta} \leq \frac{m_{\gamma}c^{2}}{2}.
\end{equation}
For the $P_{\zeta}<0$ case, Eq. (\ref{eq:poper}) can be written as
\begin{equation}
P_{\mu}P^{\mu}-m_{e\!f\!f}^{2}c^{2}= 0,
\end{equation}
with
\begin{equation}
m_{e\!f\!f}^{2}=m^{2}+\frac{2m_{\gamma}|P_{\zeta}|}{c^{2}}.
\end{equation}
Thus, the mass of the particle gets modified.\\

It can be shown that the canonical Hamiltonian is zero and thus the action
of the system in terms of the phase-space variables is
\begin{eqnarray}
 S&=& \frac{1}{2}\int d\tau \Bigg[P_{\zeta} \dot \zeta +P_{\mu}\dot X^{\mu}
\nonumber\\
& & -\lambda \left(P_{\zeta}+\frac{1}{2m_{\gamma}}\left(P_{\mu}P^{\mu}
-m^{2}c^{2}\right)\right)\Bigg].
\end{eqnarray}
Notice that, as $\phi$ is the only constraint, it is of first class
\cite{dirac:gnus}. Then, according to Dirac's method, the physical
states of the quantum system are those satisfying
\begin{equation}
\left[\hat P_{\zeta}+
\frac{1}{2m_{\gamma}}\left(\hat P_{\mu} \hat P^{\mu}-m^{2}c^{2}\right)
\right] |\psi \rangle=0. \label{eq:KGG}
\end{equation}
By assigning operators as
\begin{equation}
 \hat P_{\mu}=-i\hbar  \partial _{\mu}\qquad {\rm  and}\qquad
\hat P_{\zeta}=-i\hbar
\partial _{\zeta},
\end{equation}
the Eq. (\ref{eq:KGG}) is a generalization of the Klein-Gordon equation. An
interesting property of this is that integration of its propagator
over $\zeta$ yields the usual propagator from the Klein-Gordon
equation \cite{fey:gnus}, in consequence the usual Klein-Gordon is
an effective version of (\ref{eq:KGG}). Notice that defining the
mass operator as $\hat m^{2}=m^{2}-2m_{\gamma} \hat
P_{\zeta}/c^{2},$ the Eq. (\ref{eq:KGG}) takes the form
\begin{equation}
\left[\hat P_{\mu} \hat P^{\mu}-\hat m^{2}c^{2} \right] |\psi
\rangle=0. \label{eq:KGG1}
\end{equation}
We can consider that this Klein-Gordon equation corresponds to a
particle whose mass depends on the physical state. A more detail
analysis of this equation can be found in \cite{Fanchi:gnus}.\\

Eq. (\ref{eq:KGG}) was originally proposed by V.~Fock
\cite{fock:gnus} and was later considered by Stueckelberg and Nambu
\cite{St:gnus}. For the $m\not= 0$ case, a derivation of action
(\ref{eq:accredrep}) can be found in \cite{david:gnus}. In the next
subsection we will use the action (\ref{eq:accredrep}) to obtain a
realization of the Snyder-like noncommutative spaces.

\subsection{Snyder space-time}
\label{ss:snyder}

The action (\ref{eq:accredrep}) is invariant under
reparametrizations that implies that the system has gauge freedom.
The arbitrariness of the gauge is essentially the freedom to choose
the time. Let us now see what happens by fixing a gauge on this
system. It can be shown that by imposing
\begin{equation}\label{con2}
\chi=\zeta-\tau\approx 0,
\end{equation}
the equations of motion (\ref{eq:eq}) can be written as
\begin{eqnarray}
\ddot X_{\mu} &\!\!=\!\!& 0,\\
\dot X_{\mu} \dot X^{\mu} &\!\!=\!\!& lc^{2}, \qquad l={\rm const}.
\label{eq:rest}
\end{eqnarray}
For $l=0$, these are the equations of motion of a massless relativistic
particle. For $l=1$, they are those of a relativistic particle with mass; and
for $l=-1$, they are the equations of motion of a tachyon.\\

Considering now the gauge condition
\begin{equation}
\chi_{1}=A\tau+ B\zeta+C\zeta P_{\zeta}+ X^{\mu}P_{\mu},
\qquad A,B,C={\rm const},
\label{eq:condicion}
\end{equation}
which is an appropriate choice as by defining $\chi_{2}=\phi$, one gets
\begin{eqnarray}
\{\chi_{1},\chi_{2}\}&=&B+
CP_{\zeta}+\frac{1}{m_{\gamma}}P_{\mu}P^{\mu}\nonumber\\
&=&B+\frac{P_{\mu}P^{\mu}}{m_{\gamma}}\left(1-\frac{C}{2}\right)+
C\frac{mc^{2}}{2}\not =0
\end{eqnarray}
and therefore $(\chi_1,\chi_2)$ forms a second-class constraint set
\cite{dirac:gnus}. Thus, matrix $C_{\alpha\beta}=
\{\chi_{\alpha},\chi_{\beta}\}$ and its inverse $C^{\alpha\beta}$
are well defined. Notice that this gauge remains valid even for the
$m=0$ case. Now, by fixing the gauge,  the  constraint
(\ref{eq:poper}) and the gauge condition (\ref{eq:condicion}) are a
pair of second class constraints. In consequence, we need to change
the Poisson's into Dirac's brackets \cite{dirac:gnus}. If $F$ and
$G$ are functions from phase space, Dirac's brackets are defined as
\begin{equation}
\{F,G\}^{*}=\{F,G\}-\{F,\chi_{\alpha}\}
C^{\alpha\beta}\{\chi_{\beta},G\}.
\end{equation}
In particular, for $X_{\mu}$ and $P_{\nu}$ one obtains
\begin{eqnarray}
\{X_{\mu},X_{\nu}\}^{*} &\!\!=\!\!& -\frac{d}{\hbar^{2}}L_{\mu\nu},\quad
L_{\mu\nu}=X_{\mu}P_{\nu}-X_{\nu} P_{\mu},\ \ \ \ \label{eq:d1}\\
\{X_{\mu},P_{\nu}\}^{*} &\!\!=\!\!& \eta_{\mu\nu}
-\frac{d}{\hbar^{2}}P_{\mu}P_{\nu},\label{cxp}\\
\{P_{\mu},P_{\nu}\}^{*} &\!\!=\!\!& 0\label{eq:d3},
\end{eqnarray}
where
\begin{equation}
d=\frac{\hbar^{2}}{Bm_{\gamma}+\left(1-\frac{C}{2}\right)
P_{\mu}P^{\mu}+\frac{C}{2}m^{2}c^{2}}.\label{eq:a}
\end{equation}
To quantize this system we promote the Dirac's brackets to
commutators and then by quantizing this theory within the canonical
formalism one obtains a noncommutative space-time. Notice that $d$
depends on the momentum $P_{\mu}$, and this in principle appears to
imply a problem of ordering in the commutation rules. However, as
\begin{equation}
\{P_{\alpha}P^{\alpha}, L_{\mu\nu}\}^{*}
=\{P_{\alpha}P^{\alpha},P_{\mu}P_{\nu}\}^{*}=0,
\end{equation}
the problem does not actually exist.\\

Now, if $C=2$ then $d$ is a constant. By taking $C=2$ and
$B=-2m_{\gamma}c^{2}$, then $d=-a^{2}$ is negative for both the particle
with or without mass. In such a case the Dirac brackets
(\ref{eq:d1})--(\ref{eq:d3}) become
\begin{eqnarray}
\{X_{\mu},X_{\nu}\}^{*} &\!\!=\!\!& \frac{a^{2}}{\hbar^{2}}
L_{\mu\nu},\ \ \ \ \ \\
\{X_{\mu},P_{\nu}\}^{*} &\!\!=\!\!& \eta_{\mu\nu}+\frac{a^{2}}{\hbar^{2}}
P_{\mu}P_{\nu}, \\\
\{P_{\mu},P_{\nu}\}^{*} &\!\!=\!\!& 0,
\end{eqnarray}
and so a realization of the noncommutative space-time defined by the
commutation rules (\ref{eq:1})--(\ref{eq:3}) holds. For this the
time is quantized in units of
\begin{equation}
\frac{a}{c}=\frac{\hbar}{c^{2}\sqrt{2m_{\gamma}^{2}-m^{2}}}.
\end{equation}
On the other hand, if $C=2$ and $B=m_{\gamma}c^{2}$ then $d>0$ and the SST
holds. In this case the space gets discretized \cite{sn:gnus} in quanta of
\begin{equation}
\sqrt{d}= a =\frac{\hbar}{\sqrt{m_{\gamma}^{2}c^{2}+m^{2}c^{2}}}.
\end{equation}
For $m\not= 0$ this length-scale is proportional to Compton's length; just
as Snyder conjectured \cite{sn:gnus}. Some realizations of the SST lose
meaning in the $m=0$ case \cite{girelli:gnus}, but notice that this does
not happen in this model.\\

\subsection{Boundary conditions and general action}
\label{ss:cond}
Boundary conditions are an important element to quantize a system
\cite{he-dav:gnus}. For this reason we look for boundary conditions
consistent with this system.\\

Clearly in this case it is not possible to fix variables $X^{\mu}$
at the boundary because they do not commute. It can be shown from Eqs.
(\ref{eq:d1})--(\ref{eq:d3}) that $\{P_\zeta,P_\mu\}^{*}=0$. Thus, $(P_\zeta,P_\mu)$
forms a complete set of commuting variables, which indicates that these can be
fixed at the action boundaries. The corresponding action in this case is
\begin{eqnarray}
S_{sp}&=&\int^{\tau_2}_{\tau_1} d\tau \Bigg(-\zeta \dot P_\zeta-X^\mu
\dot P_\mu \nonumber\\
 & &-\lambda \left(P_{\zeta}+\frac{1}{2m_{\gamma}}\left(P_{\mu}P^{\mu}
-m^{2}c^{2}\right)\right)\Bigg) \label{act7}.
\end{eqnarray}
By introducing the constraints $\chi_{1}$ and $\chi_{2}$ into this, one gets to
\begin{equation}
S_{rsp}=-\int^{\tau_2}_{\tau_1} d\tau
\left(X^{\mu}+\frac{P^{\mu}}{m_{\gamma}}
\left(\frac{A+X^{\alpha}P_{\alpha}}{B-Ch}\right)
\right) \dot P_{\mu}, \label{eq:momentc}
\end{equation}
with $h=\frac{1}{2m_{\gamma}}(P_{\mu}P^{\mu}-m^{2}c^{2})$. For this action the
boundary conditions are
\begin{equation}\label{bc8}
P_\mu(\tau_1)=P_{\mu 1}, \ \ \ \ P_\mu(\tau_2)=P_{\mu 2}.
\end{equation}
Notice that by taking $A=0$ in the constraint $\chi_{1}$, Dirac's brackets
remain unchanged. Considering this and using (\ref{eq:a}) we can define
\begin{eqnarray}
g^{\alpha\beta}&=&\eta^{\alpha\beta}+\frac{P^{\alpha}P^{\beta}}
{m_{\gamma}(B-Ch)}\nonumber
=\eta^{\alpha\beta}+\frac{P^{\alpha}P^{\beta}d}{\hbar^{2}-
P_{\mu}P^{\mu}d},
\end{eqnarray}
with which (\ref{eq:momentc}) becomes
\begin{equation}
S_{rsp}=-\int^{\tau_2}_{\tau_1} d\tau g^{\alpha\beta}(P)
X_{\alpha}\dot P_{\beta}.
\end{equation}
This is an action with a metric depending on the momenta. Now, the Hamiltonian
action of a system in a curved space-time with metric $G_{\alpha\beta}(X)$
and Hamiltonian $H$ can be written in the form
\begin{equation}
S=\int^{\tau_2}_{\tau_1} \left( d\tau G_{\alpha\beta}(X)
\dot X^{\alpha}P^{\beta}-H(X,P) \right).
\end{equation}
By analogy we propose the action of a particle in the Snyder-like
 space-time and Hamiltonian $H$ as
\begin{equation}
S_{S}=\int^{\tau_2}_{\tau_1} d\tau \left(-g^{\alpha\beta}(P)
X_{\alpha} \dot P_{\beta} -H(X,P)\right).
\end{equation}
By a direct calculation can be shown the the dynamics produced by
this action is consistent with the symplectic structure
(\ref{eq:d1})--(\ref{eq:d3}) and so the quantum version of this
system has the Snyder-like space-time
as its background.\\

Another interesting point of the reduced action (\ref{eq:momentc}) is that by
choosing
\begin{eqnarray}
\tilde X^\alpha &=& g^{\alpha\beta} X_\beta \\
\tilde P_\alpha &=& P_\alpha,
\end{eqnarray}
as a new set of phase-space coordinates, this corresponds to a local
Darboux map \cite{Marsden:gnus} that transforms the symplectic structure of
Snyder-like
(\ref{eq:d1})--(\ref{eq:d3}) into the usual one. Hence,
in this set of coordinates one obtains the classical dynamics of an
ordinary particle. We point out, however, that this map is not canonical
and therefore quantum theory will not be equivalent.\\

\section{Summary}
\label{s:summ} We have presented three space-times discrete in time
which are compatible with the Lorentz symmetry, with two of these
spaces also compatible with the invariance under translations. It
was shown that all these spaces are noncommutative, one of them has
commutation rules similar to the SST and the other two similar to
the Yang space. Moreover, by using a parametrized relativistic
particle we obtain a realization of the Snyder-like spaces. Contrary
to other realizations reported, the SST realization remains
meaningful even the for the massless particle. Finally, a general
form for the action of a particle in this kind of noncommutative
spaces with arbitrary Hamiltonian is proposed.\\

\section*{Acknowledgments}
\label{s:ack} One of us (J.D.V.) acknowledges partial support from
SEP-CONACYT project 47211-F and DGAPA-UNAM grants IN104503 and
IN109107.

\end{document}